\documentclass[utf8]{frontiersSCNS} 

\usepackage{url,hyperref,lineno,microtype,subcaption}
\usepackage[onehalfspacing]{setspace}
\usepackage[utf8]{inputenc}
\usepackage{CJKutf8}
\usepackage{arydshln}
\usepackage{array}
\newcolumntype{C}[1]{>{\centering\arraybackslash}p{#1}}
\newcolumntype{R}[1]{>{\raggedleft\arraybackslash}p{#1}}

\def\cjkja#1{\begin{CJK}{UTF8}{min}#1\end{CJK}}
\linenumbers

\def\keyFont{\fontsize{8}{11}\helveticabold }
\def\firstAuthorLast{Tran {et~al.}} 
\def\Authors{Vu Tran\,$^{1,*}$ and Tomoko Matsui\,$^{1,2}$}
\def\Address{$^{1}$ Risk Analysis Research Center, The Institute of Statistical Mathematics,  Tokyo, Japan \\
$^{2}$ Department of Statistical Modeling, The Institute of Statistical Mathematics, Tokyo, Japan}
\def\corrAuthor{Vu Tran}

\def\corrEmail{vutran@ism.ac.jp}

\begin{document}
\nolinenumbers
\onecolumn
\firstpage{1}

\title[COVID-19 Tweet Analysis]{Tweet Analysis for Enhancement of COVID-19 Epidemic Simulation: A Case Study in Japan} 

\author[\firstAuthorLast ]{\Authors} 
\address{} 
\correspondance{} 

\extraAuth{}

\maketitle

\begin{abstract}
\section{}
The COVID-19 pandemic, which began in December 2019, progressed in a complicated manner and thus caused problems worldwide. Seeking clues to the reasons for the complicated progression is necessary but challenging in the fight against the pandemic. We sought clues by investigating the relationship between reactions on social media and the COVID-19 epidemic in Japan. Twitter was selected as the social media platform for study because it has a large user base in Japan and because it quickly propagates short topic-focused messages (“tweets”). Analysis using Japanese Twitter data suggests that reactions on social media and the progression of the COVID-19 pandemic may have a close relationship. Experiments to evaluate the potential of using tweets to support the prediction of how an epidemic will progress demonstrated the value of using epidemic-related social media data. Our findings provide insights into the relationship between user reactions on social media, particularly Twitter, and epidemic progression, which can be used to fight pandemics. 

\tiny
\keyFont{\section{Keywords:} COVID-19, SEIR model, Simulation, SNS, Twitter, Emotion, Emoji} 
\end{abstract}

\section{Introduction}
We investigated the potential of using data from social media to enhance the prediction and simulation of an epidemic’s* progression. A case study was carried out using Twitter data related to the COVID-19 epidemic in Japan. The COVID-19 pandemic has been causing global problems that have affected everyone for a lengthy period, and the end is not in sight. During the pandemic, people are seeking information or clues for use in deciding their next actions through a variety of channels: newspapers, TV, and especially social media.

Studies have shown that social media greatly affects society. Twitter is one of the largest social media platforms worldwide that greatly affects several aspects of society (daily life conversations, news reports, event advertisements, etc.) in various domains (health, entertainment, economics, research, politics, etc.). During the COVID-19 pandemic, a large volume of information on Twitter regarding the infection situation, symptoms, treatment, vaccinations, restrictions, and so on is being continuously shared and discussed. Users can share their emotions and opinions regarding the information instantaneously without geographical limitations. The effects of these emotions and opinions can thus spread rapidly.

Research on predicting the progression of the COVID-19 pandemic has received much attention worldwide. Early prediction is important for implementing countermeasures against its spread. Epidemiological models, e.g., the susceptible-exposed-infected-recovered (SEIR) model, are commonly used for such prediction. The parameters are obtained from observed data or set on the basis of predefined scenarios. Complex problems, e.g., the emergence of new variants, diverging government policies, and diverging public perceptions, have arisen as the pandemic has lasted longer and longer. Many countries, including Japan, have already experienced more than four waves of the pandemic. To tackle the complicated progression of the COVID-19 pandemic and to deal with the challenge of obtaining parameters reflecting reality as conditions continue to change, recent research has focused on utilizing extra information to enhance the prediction model. 

One way to obtain such information is to monitor social media: Twitter, Facebook, Reddit, etc. Social networking services, which were initially simply playgrounds for small communities of computer users, have evolved into large social media platforms connecting both online and offline social networks. Twitter, one of the largest social media platforms, has been targeted in numerous studies aimed at identifying the personality traits of social media users~\citep{wald2012using,sumner2012predicting}.  Monitoring social media is an attractive approach to gathering data for use in various types of research~\citep{azzaoui2021sns,yoneoka2020early,alessa2019preliminary,yoo2020predictors}. Several epidemic-related behaviors can be observed on social media, for instance, health information seeking. A heavy reliance on social media has been observed during the COVID-19 pandemic ~\citep{neely2021health,dadaczynski2021digital,skarpa2021information}. Several studies of the formation of pandemic waves have revealed an association between non-pharmaceutical interventions and social behaviors \citep{cacciapaglia2021multiwave,kupferschmidt2020can,ravi2021can}. Previous work on using Twitter data to support predicting of COVID-19 epidemic progression have used tweet counts (with relevant keywords)~\citep{yousefinaghani2021prediction} and tweet full-text analysis~\citep{azzaoui2021sns}. 

In research regarding social media affecting social behavior, emotion is a critical aspect~\citep{settanni2015sharing,park2012depressive,wald2012using}. Van Bavel et al. observed that, especially in the current COVID-19 pandemic, "Social networks can amplify the spread of behaviors that are both harmful and beneficial during an epidemic, and these effects may spread through the network to friends, friends’ friends and even friends’ friends’ friends" \citep{van2020using}. The social network created by a popular social media platform such as Twitter is huge with instant connectivity without geographical limitations. This means that popular social media platforms can amplify the spread of behaviors to a magnitude much much greater than offline social networks (e.g., neighborhoods). Several studies have revealed the emotions of social media users towards COVID-19 progression \citep{wheaton2021fear,arora2021role,toriumi2020social,dyer2020public,mathur2020emotional,kaur2020monitoring}. 

We have investigated the utilization of emoji usage on Twitter to capture changes in the emotions of social media users for use in enhancing epidemiological models. Several studies have focused on capturing emotion from texts including posts on Twitter (“tweets”). However, accurately understanding emotional tweets by using full-text analysis is a challenging task. Emoji analysis is an attractive approach because social media users tend to express emotions using non-verbal communication, and they share a common understanding of many emoji.

\begin{figure}
\centering
\includegraphics[width=0.8\textwidth,trim=4cm 4cm 4cm 5cm,clip]{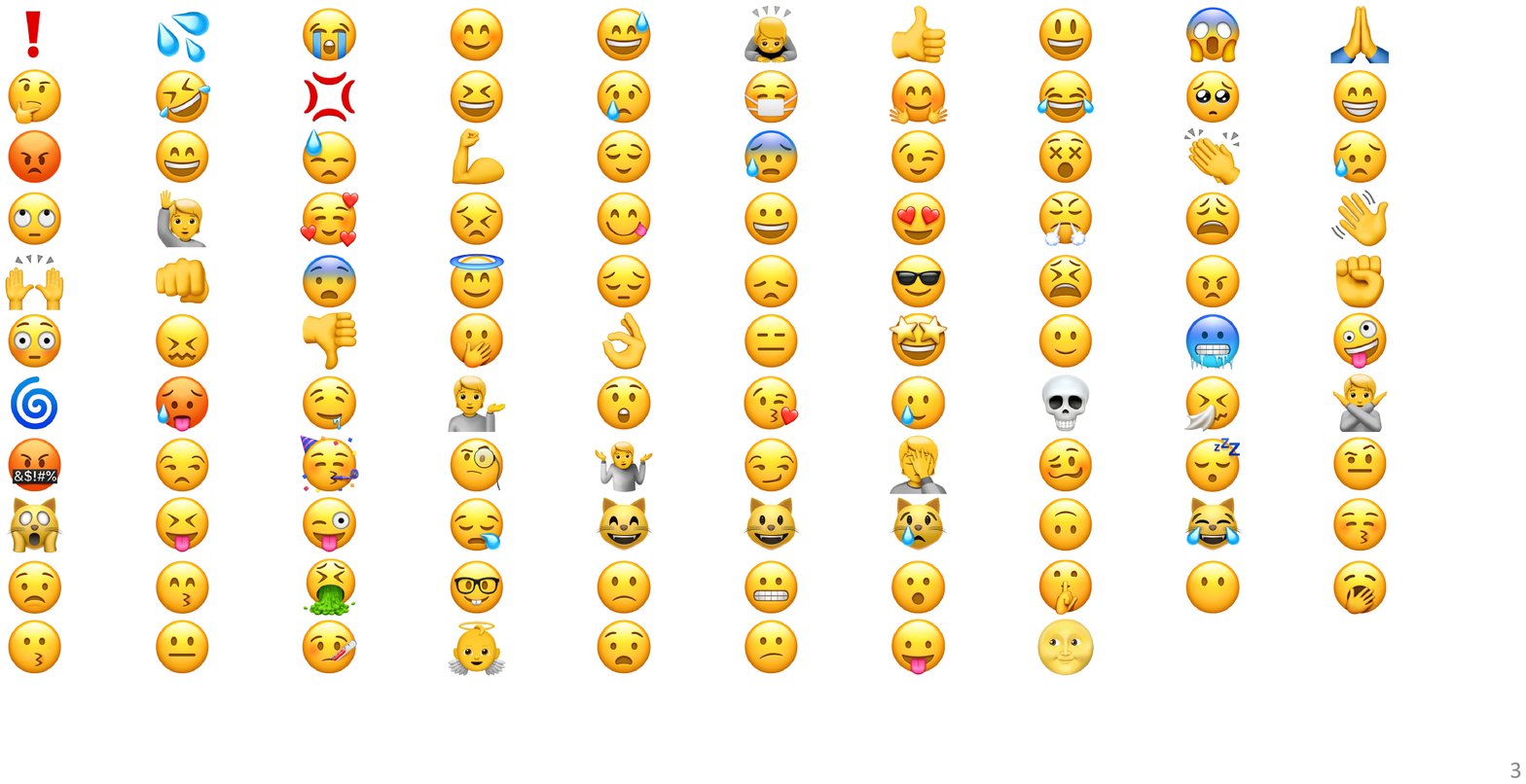}
\caption{Commonly used emoji in COVID-19 related tweets.}
\label{fig:emoji-list}
\end{figure}

Several studies have shown that emojis are used on social media as non-verbal communication cues to assist communication~\citep{suntwal2021pictographs,elder2018words,cheng2017mean,lo2008nonverbal}. Emoji are digital images depicting simple illustrations including facial expressions (smiley face, crying face, fearful face, scary face, etc.), as illustrated in Figure~\ref{fig:emoji-list}. Emotional messages can be directly expressed through emoji. Because social media users share a common understanding of many emoji, emotions can be effectively and conveniently communicated through emoji. 

One crucial point when using social media data, particularly Twitter data, is that social media users may become less engaged, i.e., performing fewer actions such as ``liking," ``commenting," and ``sharing," as the pandemic lasts longer and longer\citep{yousefinaghani2021prediction}. When engagement drops to a certain level, social media data becomes less representative of behavioral changes. The results of a study using Twitter data from the U.S. and Canada by \cite{yousefinaghani2021prediction} suggest that there will be less engagement through social media due to a feeling of exhaustion as waves of the pandemic continue. In this study, we also took into consideration the results of previous studies using Japanese Twitter data. 

\section{Materials and Methods}
\subsection{Data Collection}
The data consisted of tweet counts and COVID-19 infection data from Japan. 

\begin{figure}[h!]
\centering
\includegraphics[width=.8\textwidth,trim=5cm 4cm 5cm 4cm,clip]{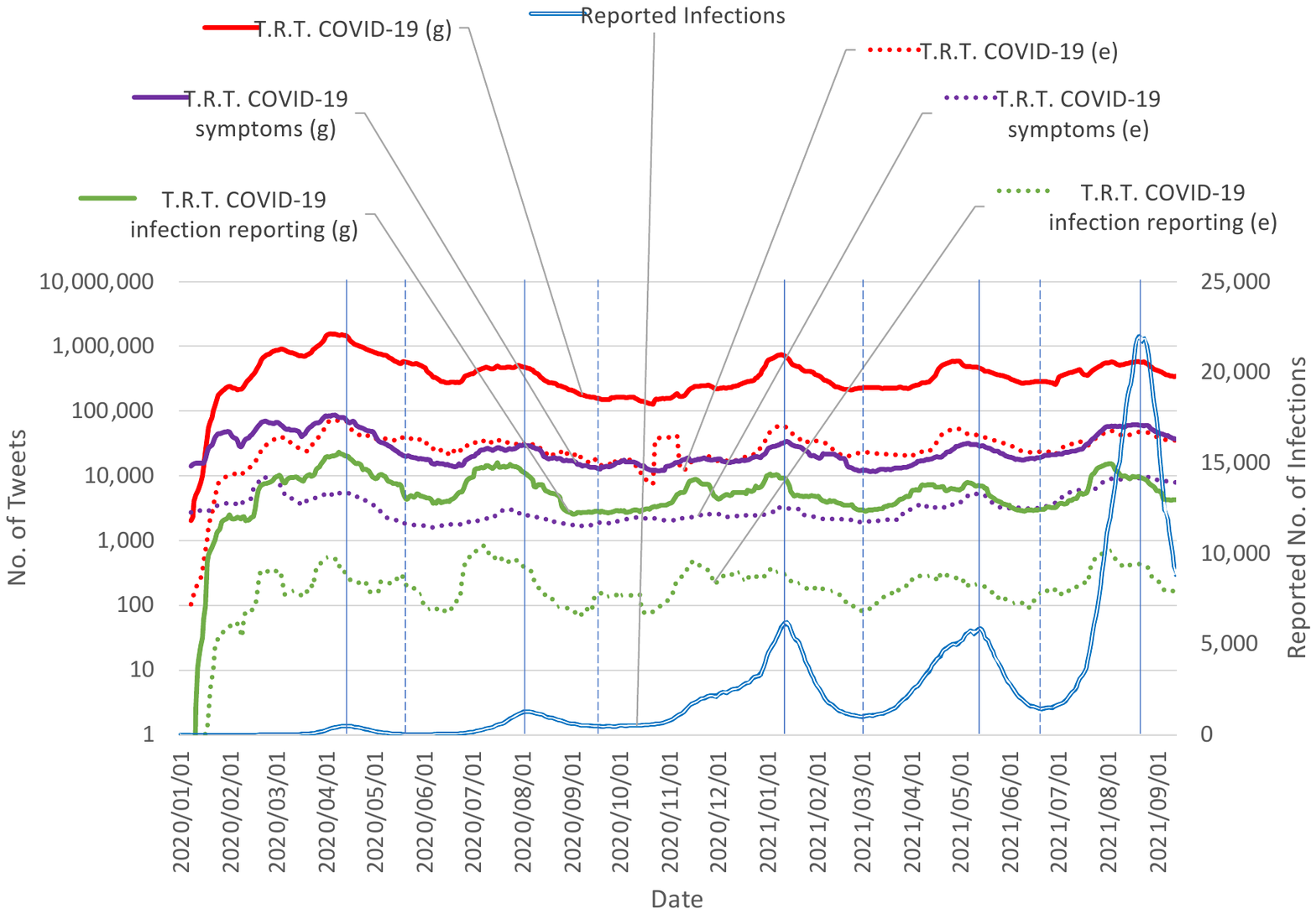}
\caption{Daily chart of tweet counts vs. reported COVID-19 infections in Japan (values were smoothed by 15-day moving average). The data suggest that the number of COVID-19 related tweets has been correlated to some degree with the progression of the epidemic in Japan since the beginning of the epidemic. (\textit{T.R.T.: Tweets related to})}
\label{fig:tweetcount}
\end{figure}

The tweet count data were collected using the Twitter API (version 2) with academic research access. Several settings were considered, from the general COVID-19 related tweet count to more fine-grained target subsets of keywords. Three sets of keywords were used: COVID-19 related set, COVID-19 symptom related set\footnote{The symptom-related keywords were obtained from \url{https://www.kansensho.or.jp/ref/d77.html} and \cite{yousefinaghani2021prediction}}, and COVID-19 infection reporting related set. For each set, the collections were further filtered to retain only tweets containing emojis. The COVID-19 related set was the primary set used. The other sets were used for an ablation study and analysis of the characteristics of the tweets. The details of the settings are shown in Table~\ref{tab:tweet-stats}. The collected data show that the number of COVID-19 related tweets has been correlated to some degree with the COVID-19 epidemic progression since the beginning of the epidemic (Figure ~\ref{fig:tweetcount}). 

\begin{table}[]
\caption{Tweet count settings.}
\centering
\renewcommand{\arraystretch}{1.5}
\begin{tabular}{p{2.5cm} C{1.5cm} p{10.5cm} R{1.2cm}}
\hline \hline
\textbf{Tweets Related To} & \textbf{Only Tweets with Emojis} & \textbf{Query Keywords } & \textbf{Daily No. of Tweets} \\ \hline \hline
COVID-19 (g)  & No & \cjkja{新型コロナ , コロナ感染 , コロナ禍 , コロナワクチン , 緊急事態宣言 , まん延防止 , 感染者} 
\par
(\textit{translation: [new-variant corona, corona infection, corona disaster, corona vaccine, emergency declaration, spread prevention, infected person/people]})
& 414,576 \\ \hdashline
COVID-19 (e) & Yes & same as above & 29,484 \\ \hline
COVID-19 symptoms (g)  & No & \cjkja{発熱 , 鼻汁 , 咽頭痛 , 咳嗽 , 嗅覚異常 , 味覚異常 , 息切れ , 咳 , のどの痛み , 喉の痛み , 嗅覚障害 , 味覚障害},
\par \textbf{excluding} \{\cjkja{風邪 , インフルエンザ , 糖尿病 , マラリア , サタデーナイトフィーバ , 喫煙 , たばこ , アレルギー , アレルギ} \} 
\par

(\textit{translation: [fever, nasal discharge, sore throat, cough, dysosmia, dysgeusia, shortness of breath, cough, sore throat, sore throat, dysosmia, dysgeusia]},
\par
\textit{excluding: \{cold, influenza, diabetes, malaria, Saturday night fever (a movie-related reference to the risk of going out dancing), smoking, tobacco, allergies, allergies \}})
& 28,814 \\ \hdashline
COVID-19 symptoms (e) & Yes & same as above & 3,597 \\ \hline
COVID-19 infection reporting (g) & No & \cjkja{感染者数 , 陽性者数} 
\par
(\textit{translation: [number of infected people, number of confirmed positive cases]}) & 6,518
\\ \hdashline
COVID-19 infection reporting (e) & Yes & same as above & 232 \\ \hline \hline

\end{tabular}

\label{tab:tweet-stats}
\end{table}

The COVID-19 infection reporting data for Japan were obtained from JX Press\footnote{https://jxpress.net/}. The dataset contains daily infection reports for all prefectures in Japan. It was used for training or calibrating two core models used by the epidemic simulation system described in Subsections~\ref{sec:snslstm} and~\ref{sec:simusys}.

\begin{figure}[h!]
\centering
\includegraphics[width=0.75\textwidth,trim=5cm 5cm 8cm 4cm,clip]{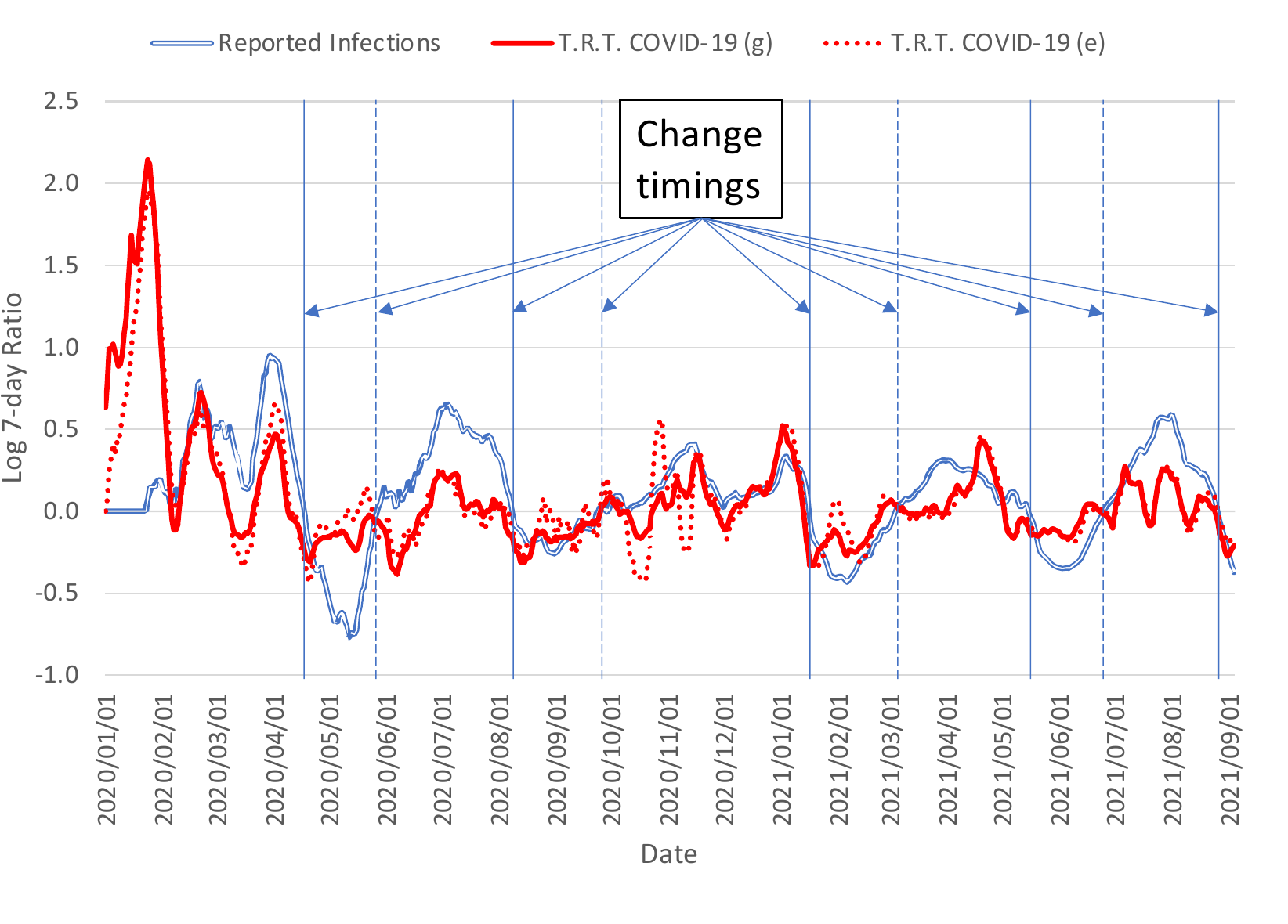}
\caption{Logarithm of increasing rate of the day of the week for reported infections and tweet counts calculated using Equation~\ref{eq:7day-ratio-log}. (\textit{T.R.T.: Tweets related to})}
\label{fig:transformed-data}
\end{figure}

\begin{figure}[h!]
\centering
\includegraphics[width=1\textwidth,trim=5cm 3cm 2cm 2cm,clip]{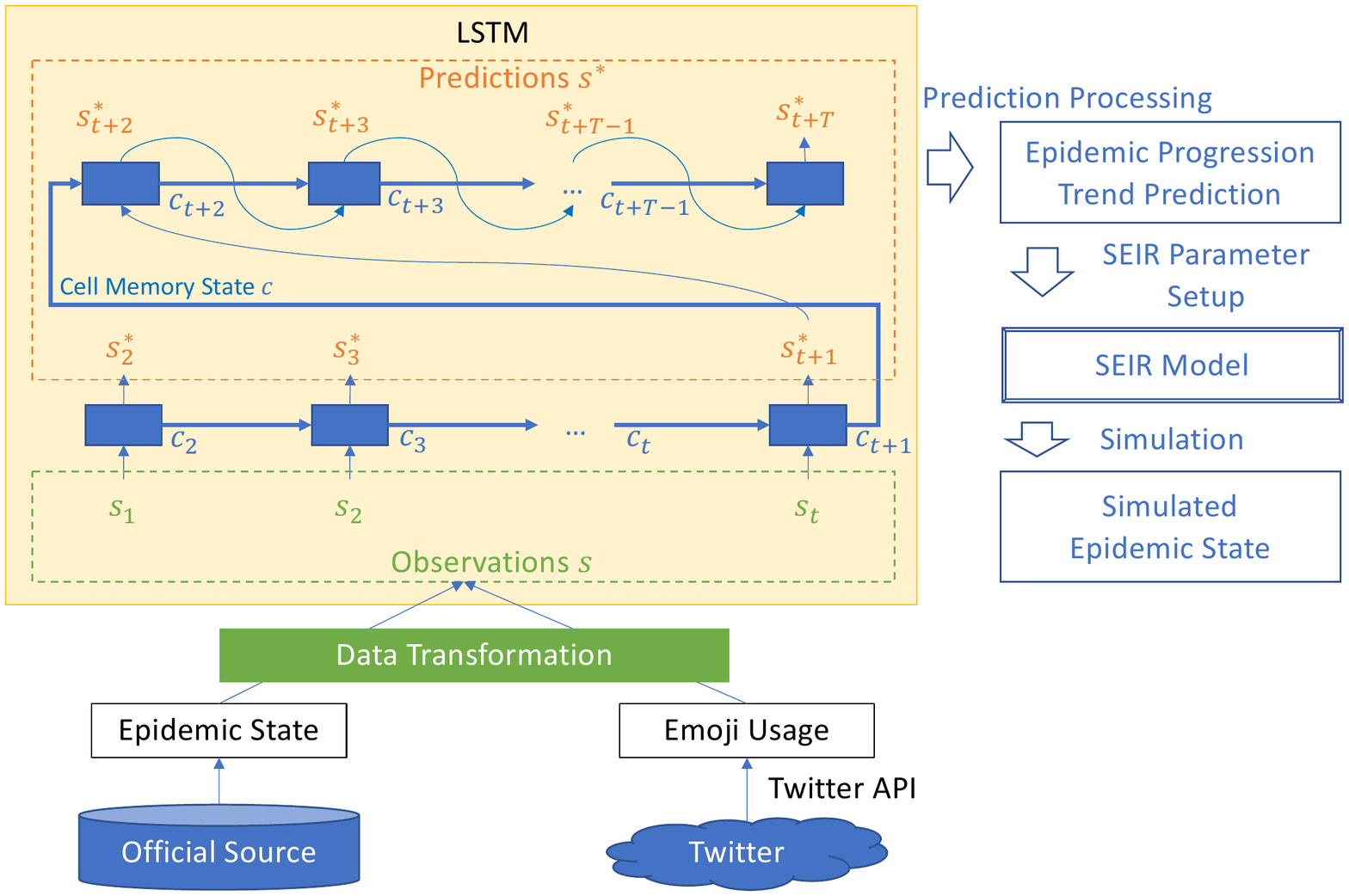}
\caption{COVID-19 epidemic simulation system ($t$ marks end timing of observable data).}
\label{fig:simu-sys}
\end{figure}

\subsection{SNS Reaction Trend \& COVID-19 Epidemic Progression Change Prediction}
\label{sec:snslstm}
As shown in Figures~\ref{fig:tweetcount} and~\ref{fig:transformed-data}, the trend in reported infections or cases was similar to the trend in the reaction level on social media. This suggests a non-negligible correlation between the two signals. Predicting the trend of changes in the epidemic progression helps to set up appropriate  scenarios for simulating the future epidemic state, which in turn supports policy makers. In this sense, given the suggestion of a potential relationship between the trends of the two signals, additional information from social media reactions may further support predicting changes in the epidemic progression. 

Here, the trend representations were estimated using the ratio of the signals for days $t$ and $t-7$, which were the same day of the week:
\begin{equation}\label{eq:7day-ratio-log}
s_t = log(\frac{o_t}{o_{t-7}}),      
\end{equation}
where $o_t$ represents the two signals, the reactions on Twitter measured by tweet count and the epidemic state estimated from the reported number of new infections on day $t$, and $s_t$ represents the trend measured as the 7-day change. This transformation absorbs the weekly effect observed in the Japanese data. The transformation was further smoothed by a 15-day moving average. 

To model the relationship between the trend in social media reactions and the trend in epidemic progression, we utilized a long short-term memory (LSTM) neural network~\citep{hochreiter1997long}, a well-known and successful neural network architecture in time-series modeling, and the multivariate time-series of the two trends. LSTM neural networks have been used in various domains for modeling time-series and have achieved practical results. In previous studies of COVID-19 epidemic prediction systems, LSTM models were used as the core models~\citep{chimmula2020time,shahid2020predictions,kirbacs2020comparative}. 

To cope with the unknown complexity of the relationship between the two time-series, we use an ensemble system of multi-layer LSTM models with various hyperparameter (number of layers, number of neurons) settings and parameter initialization of the LSTM models~\footnote{$ \mbox{ no. of layers} \in \{1,2,3,4\} \times \mbox{\ no. of neurons} \in \{4,8,16\} \times \mbox{\ no. of initializations} \in \{128\}$}. 

The LSTM system is optimized by minimizing the mean squared error:
    \begin{equation}
    MSE(s_{2:t},s^*_{2:t}) = \frac{1}{t-1}\sum_{k=2}^t \frac{1}{d} \sum_{j=1}^d   {(s_{k,j} - s^*_{k,j})^2},
    \end{equation}
where $t$ marks the end of the observable or training data, $d=2$ is the number of time-series (including the trend of reactions on Twitter and the trend of the epidemic progression), and $s,s^*$ are the observed data and the corresponding predictions. 

The inference procedure has two phases. In the first phase, the LSTM ensemble system receives observed data $\{s_k | k \in [1,t]\}$ up to time $t$ and uses them to create memory state $c_{t+1}$ and prediction $s^*_{t+1}$ (Equation~\ref{eq:lstm-p1}). In the second phase, from input time-step $t+1$, the prediction of the previous time-step is used as the input to predict the next time-step (Equation~\ref{eq:lstm-p2}). The inference procedure is illustrated in the ``LSTM" box at the top-left of Figure~\ref{fig:simu-sys}. In the training or optimization process, only the first phase is invoked, and predictions $s^{*}_{2:t} = \{s^{*}_k | k \in [2,t]\}$ are used for the aforementioned optimization. 
\begin{align}
    \{s^{*}_{k+1}, c_{k+1}\} &= \mbox{LSTM}(s_{k},c_{k}) \mbox{ for } k \in [1, t] \label{eq:lstm-p1} \\
    \{s^{*}_{k+1}, c_{k+1}\} &= \mbox{LSTM}(s^{*}_{k},c_{k}) \mbox{ for } k \in [t+1,t+T-1], \label{eq:lstm-p2}
\end{align}
where $k$ is the input time-step, $t$ marks the end of the observable data, $T$ is the length of the prediction period, $c$ is the memory state of the LSTM, and $s,s^*$ are the observed data and corresponding predictions. 

The outputs of the \textit{change prediction} model are used for setting up the COVID-19 simulation system described in the next subsection. The outputs of the \textit{change prediction} model are processed to identify the timings when the predicted values change sign (illustrated in Figure~\ref{fig:transformed-data}): 
\begin{itemize}
\item From positive to negative: the signal progression changes from increasing (up-trend) to decreasing (down-trend).
\item From negative to positive: the signal progression changes from decreasing (down-trend) to increasing (up-trend).
\end{itemize}

\subsection{COVID-19 Epidemic Simulation System}
\label{sec:simusys}
The COVID-19 epidemic simulation system consists of two stages: 1) \textit{change prediction}, 2) \textit{simulation}. The \textit{change prediction} is executed as described in Subsection~\ref{sec:simusys}. The \textit{simulation} is executed using SEIR, a common epidemic model. The overall flow of the system illustrated in Figure~\ref{fig:simu-sys} is as follows.
\begin{enumerate}
\item \textit{Data collection}: collect tweet count and COVID-19 epidemic state;
\item \textit{Data transformation}: estimate trend representations for tweet count and COVID-19 epidemic progression;
\item \textit{\underline{Change prediction}}: predict trends and identify change timings;
\item \textit{SEIR model parameter setup}: set SEIR model parameters in accordance with the identified change timings;
\item \textit{\underline{Simulation}}: perform epidemic simulation.
\end{enumerate}

We used the simulation system proposed by \cite{lemaitre2021scenario} with a stochastic SEIR model to model the disease dynamics. The system supports multi-location epidemic modeling to estimate the \textit{force of infection} (rate at which susceptible individuals are infected) by using inter-location mobility. The formulation of the SEIR model is described in the Appendix~\ref{appe:SEIR}. We performed prefecture-wide multi-location setup. The SEIR model uses the following parameters: the latent period $\frac{1}{\sigma}$, which is the time interval between when an individual becomes infected and when he or she becomes infectious, the infectious period $\frac{1}{\gamma}$, which is the time interval during which an individual is infectious, and the effective reproduction number $R_{i}$(t) for each location $i$ at time $t$, which is the number of cases generated in the current state of a population.

While the latent period $\frac{1}{\sigma}$ and infectious period $\frac{1}{\gamma}$ depend on the COVID-19 variant, the effective reproduction number $R_{i}(t)$ depends not only on the variant but also on the contact rate in the community, which changes as the behaviors of the community members change. During one wave of the COVID-19 epidemic, the change in $R_{i}(t)$ was greatly affected by behavioral changes due to perceived events, e.g., surging of cases and policy changes (emergency declarations), resulting in up trends and down trends in the epidemic progression. Hence, determining $R_{i}(t)$ is the key to effective simulation. 

A set $\textbf{R}_i = \{R_{i}(t)\}$ was obtained using the calibration method used by \cite{lemaitre2021scenario} for the period from 2020/12/24 to 2020/01/21 (the 3$^{rd}$ wave in Japan) using the observed epidemic data. Two subsets of $R_{i}(t)$ were established: up-trend set $\textbf{R}^u_i$ (2020/12/24 – 2020/01/06) and down-trend set $\textbf{R}^d_i$ (2021/01/07 – 2021/01/21). 

In the \textit{simulation} period from 2021/04/23 to 2021/06/30, for each trend (up or down) time span $[t_s,t_e]$, a set of $\{R_{i} (t)\}$ for each location $i$ was drawn from a uniform distribution: 
\begin{align}\label{eq:pick-Rt}
    R_{i} (t)_{| t_s \leq t \leq t_e} \sim \mathcal{U}[m^{(p)}_i,M^{(p)}_i],  
\end{align}
where $m^{(p)}_i,M^{(p)}_i$ are, respectively, the minimum and maximum values of a set of previously obtained reproduction numbers, which can be either $\textbf{R}^u_i$ or $\textbf{R}^d_i$ depending on whether time span $p$ is trending up or down. If $[t_s,t_e]$ is an up-trend time span, $\textbf{R}^u_i$ is selected, and if $[t_s,t_e]$ is a down-trend time span, $\textbf{R}^d_i$ is selected. The change timings, $t_s$ and $t_e$, are determined in the \textit{change prediction} stage, as described in Subsection~\ref{sec:snslstm}.

For evaluation, we measure the errors in the \textit{change prediction} and \textit{simulation} stages against the observed data for the period from 2021/04/23 (in the up-trend of the 4$^{th}$ wave) to 2021/06/30 (ending of the 4$^{th}$ wave). We used data from 2020/12/24 to 2020/01/21 (in the 3$^{rd}$ wave) to obtain the SEIR model parameters and data from 2020/11/15 to 2021/04/22 (the end timing of observable data) for training the \textit{change prediction} model. Two observed timings of trend changes were used for evaluation: $t_a = 2021/05/15$ and $t_b = 2021/06/25$, where $t_a$ marks the change from up-trend to down-trend, and $t_b$ marks the change from down-trend to up-trend in the epidemic progression as observed in the infection reports. 

The evaluation metric for \textit{change prediction} was the difference in days $\Delta{days}[t]$ between the predicted date $t'$ and the actual date $t$ of the trend change (Equation~\ref{eq:err-change-prediction}). 
\begin{equation}\label{eq:err-change-prediction}
    \Delta{days}[t] = t' - t
\end{equation}
The evaluation metric for \textit{simulation} was the root-mean-square error (RMSE).

\section{Results}
Table~\ref{tab:exp-main-results} shows the results for \textit{change prediction} and \textit{simulation}. Two baselines were used for reference.
\begin{itemize}
\item \textit{Baseline 1}: $R_i (t)$ was set for the entire simulation period using $\textbf{R}_i$ in the up-trend and down-trend periods of the 3$^{rd}$ wave. $R_i (t)$ were sampled for both the up-trend and down-trend periods without knowing the exact timing of the trend change.
\item \textit{Baseline 2}: $R_i (t)$ was set for the entire simulation period using $\textbf{R}^u_i$ in the up-trend period of the 3$^{rd}$ wave. $R_i (t)$ were sampled for only the up-trend period.
\end{itemize}
For our approach, we used three system settings:
\begin{itemize}
\item +\textit{change prediction} w/o using tweet data: the epidemic simulation system was setup with\textit{change prediction} using only the epidemic state data, not the tweet data.
\item +\textit{change prediction} using T.R.T. COVID-19 (g): the epidemic simulation system was setup with \textit{change prediction} using both the epidemic state data and the COVID-19 related tweet count data.
\item +\textit{change prediction} using T.R.T. COVID-19 (e): similar to setting for (g) except that tweets were filtered to remove ones not containing emoji.
\end{itemize}

The additional use of the COVID-19 related tweet count (g) resulted in better prediction of the epidemic progression trend changes than without using the count: prediction was improved by 8.5 days for $t_a$ and 6.3 days for $t_b$. This led to a reduction of $42.8\%$ in the RMSE. Given that the daily tweet count of COVID-19 related tweets filtered for emoji (e) was $92.9\%$ smaller than the more general count (g), the results are similar: the difference in \textit{change prediction} was 0.2 days for $t_a$ and 2.4 days for $t_b$, and the RMSE was $5.5\%$ worse. In all results, the predicted trend changes preceded the observed changes. The baseline results show that without estimating the trending change, the RMSE were $7.6 – 18.5$ times worse.

\begin{table}[h!]
\caption{Evaluation results for \textit{change prediction} (Equation~\ref{eq:err-change-prediction}) and \textit{simulation} (RMSE) for 4$^{th}$ wave in Japan (2021/04/23 – 2021/06/30) with two epidemic progression trend changes: $t_a = 2021/05/15$ and $t_b = 2021/06/25$. Data from 2020/12/24 to 2020/01/21 were used to obtain SEIR model parameters for up-trend and down-trend periods of COVID-19 epidemic progression. Data from 2020/11/15 to 2021/04/22 were used for training \textit{change prediction} model. (\textit{T.R.T.: Tweets related to})}
    \centering
    \renewcommand{\arraystretch}{1.2}
    \begin{tabular}{l|C{4.5cm}|R{2cm}|R{1.5cm}}
\hline \hline
\textbf{Epidemic Simulation System} & \textbf{Change Prediction} \par ($\Delta{days}[t_a]$ / $\Delta{days}[t_b]$ ) & \textbf{Simulation} (RMSE) & \textbf{Daily  No. of Tweets} \\ \hline \hline
Baseline 1 & n/a & 18,093.9 & n/a \\ 
Baseline 2 & n/a & 25,216.0 & n/a \\ \hline
\textit{+change prediction} w/o using tweet data & -16.3 / -28.0 & 2,377.9 & n/a \\ \hline
\textit{+change prediction} using T.R.T. COVID-19 (g) & \ \ -7.8 / -21.7 & 1,360.4 & 414,576 \\
\textit{+change prediction} using T.R.T. COVID-19 (e) & \ \ -8.0 / -19.3 & 1,435.1 & 29,484 \\ \hline
\end{tabular}
\label{tab:exp-main-results}
\end{table}

\section{Discussion}
The relationship between user reactions on social media and the COVID-19 epidemic progression remains close for the long term. Social media engagements related to COVID-19 have remained fairly steady over the five waves of COVID-19 epidemic surges in Japan. They reached their highest level in the first wave, dropped a bit in the second wave, and then picked up in the following waves. The engagements peaked at around the peak of each wave. This demonstrates the value of using epidemic-related social media data, particularly Twitter data. 

The 3$^{rd}$ and 4$^{th}$ waves in the period from 2020/11/15 to 2021/06/25 exhibited similar characteristics: the wave shapes were similar (Figure~\ref{fig:tweetcount}) and the vaccination rates were similar\footnote{\url{https://www.kantei.go.jp/jp/headline/kansensho/vaccine.html}}. Despite the similar wave shapes, the reactions to non-pharmaceutical interventions and emergency declarations differed between the two waves. In the 3$^{rd}$ wave, an emergency declaration was issued on 2021/01/07, and a change in the epidemic progression trend (from increasing to decreasing) was observed on 2021/01/17 (ten days later). In contrast, in the 4$^{th}$ wave, an emergency declaration was issued on 2021/04/25, and a change in the epidemic progression trend was observed on 2021/05/15 (20 days later). The 10-day later response in the 4$^{th}$ wave is attributed to reluctance to comply or exhaustion after already being subjected to two previous emergency declarations. The reluctance or exhaustion level is somewhat correlated with the reaction on social media, which was partially captured by the \textit{change prediction} model and resulted in more accurate prediction of the change in the epidemic progression trend. 

\subsection*{Future Work}
For further improvement in the \textit{simulation} results, the method for setting the SEIR model parameters needs to be further improved, especially for the setting of $R_i (t)$. In this study, the distribution from which the set of $\{R_{i} (t)\}$ for each location $i$ was drawn was assumed to be uniform, and the up- and down-trend parameter sets were manually established. The setting of the SEIR model parameters would be more challenging in periods in which the epidemic conditions greatly differed, e.g., the 5$^{th}$ wave in Japan in which the \textit{delta} variant was dominant. Viable options include selecting values from the most recent wave with adjustment for the infectious power of newer variants and selecting from the period with the most similar social media reactions although measuring similarity would be a challenging task. Furthermore, it is necessary to consider the emergence of new COVID-19 variants and how they would affect the parameters as well as the social media reactions. These challenges will be addressed in future work. 

As preparation for future work, we performed experiments on training the \textit{change prediction} model using different fine-grained tweet counts:
\begin{itemize}
\item T.R.T. COVID-19 symptoms(g)
\item T.R.T. COVID-19 symptoms(e)
\item T.R.T. COVID-19 infection reporting (g)
\item T.R.T. COVID-19 infection reporting (e).
\end{itemize}
The tweet counts are listed in Table~\ref{tab:tweet-stats}, and the results of the additional experiments are shown in Table~\ref{tab:exp-extra-results}.

\begin{table}[h!]
\caption{Tweet counts for \textit{change prediction} for 4$^{th}$ wave in Japan (2021/04/23 – 2021/06/30) with two epidemic progression trend changes: $t_a = 2021/05/15$ and $t_b = 2021/06/25$. Data from 2020/12/24 to 2020/01/21 were used to obtain SEIR model parameters for up-trend and down-trend periods of COVID-19 epidemic progression. Data from 2020/11/15 to 2021/04/22 were used for training \textit{change prediction} model. (\textit{T.R.T.: Tweets related to})}
\centering
\renewcommand{\arraystretch}{1.2}
    \begin{tabular}{l|C{4.5cm}|R{2cm}|R{1.5cm}}
    \hline \hline
\textbf{Tweet Count for \textit{Change Prediction}} & \textbf{ Change Prediction} \par ($\Delta{days}[t_a]$ / $\Delta{days}[t_b]$ ) & \textbf{Simulation} (RMSE) & \textbf{Daily No. of Tweets} \\ \hline \hline
T.R.T. COVID-19 (g) & \ \ -7.8 / -21.7 & 1,360.4 & 414,576 \\
T.R.T. COVID-19 (e) & \ \ -8.0 / -19.3 & 1,435.1 & 29,484 \\ \hline
T.R.T. COVID-19 symptoms (g) & -17.4 / -27.7 & 2,478.9 & 28,814 \\
T.R.T. COVID-19 symptoms (e) & -16.2 / -29.4 & 2,389.6 & 3,597 \\
T.R.T. COVID-19 infection reporting (g) & -13.4 / -26.4 & 2,051.2 & 6,518 \\
T.R.T. COVID-19 infection reporting (e) & -12.4 / -23.4 & 1,932.7 & 232 \\ \hline \hline
\end{tabular}
    
\label{tab:exp-extra-results}
\end{table}

Compared with using the general-topic COVID-19 related tweet counts, using more specific-topic tweet counts did not show improvement: the RMSE was $34.7\% – 82.2\%$ worse for the \textit{simulation} period. This suggests that the relationship between reactions on social media and epidemic progression is complex. The general count, covering a broad range of topics, exhibited greater predictive power than the more specific counts. Manual topic design thus may not be an efficient approach. The development of automatic topic discovery techniques for finding relevant topics discussed on social media that can support epidemic progression prediction could be promising. 

The results for tweet counts with emoji filtering (e) compared with the general tweet counts (g) showed that the emoji settings have similar representative value as the general settings: the RMSE difference was only $3.6\% – 5.8\%$ even with $87.5\% – 96.4\%$ fewer tweets. One advantage of using emoji settings is the ability to perform fine-grained analysis on specific emotions (fear, anger, etc.) represented by various emojis. Further studies on the specific emotions used by social media users for typical topics could help in discovering topics where changes in emotion could affect epidemic progression. This could be done by analyzing social media contents (emoji vs. topics) to identify emotions trending on topics relevant to epidemic progression This is left for future work. 

\appendix
\renewcommand{\thesubsection}{\Alph{subsection}}
\section*{Appendix}
\label{appe:SEIR}
In this study, we used the simulation system proposed by \cite{lemaitre2021scenario} with a stochastic SEIR model used to model the disease dynamics. This system  supports multi-location epidemic modeling to   estimate the  \textit{force of infection} using inter-location mobility. For Japan, we performed prefecture-wide multi-location setup. Given the parameters, including the reproduction numbers $R_{i}$(t), latent period $\frac{1}{\sigma}$, and infectious period $\frac{1}{\gamma}$, the transitions between the compartments  \textbf{S}usceptible, \textbf{E}xposed, \textbf{I}nfected, and \textbf{R}ecovered for each location $i$ are
\begin{align}
    N_{\mathcal{S}_i \rightarrow \mathcal{E}_i}(t) &= \mbox{Binom}(\mathcal{S}_i,1-\mbox{exp}(-\Delta t \cdot \mbox{FOI}_i (t))) \\
    N_{\mathcal{E}_i \rightarrow \mathcal{I}_i^{(1)}}(t) &= \mbox{Binom}(\mathcal{E}_i,1-\mbox{exp}(-\Delta t \cdot \sigma)) \\
    N_{\mathcal{I}_i^{(1)} \rightarrow \mathcal{I}_i^{(2)}}(t) &= \mbox{Binom}(\mathcal{I}_i^{(1)},1-\mbox{exp}(-\Delta t \cdot \gamma')) \\ 
    N_{\mathcal{I}_i^{(2)} \rightarrow \mathcal{I}_i^{(3)}}(t) &= \mbox{Binom}(\mathcal{I}_i^{(2)},1-\mbox{exp}(-\Delta t \cdot \gamma')) \\ 
    N_{\mathcal{I}_i^{(3)} \rightarrow \mathcal{R}_i}(t)       &= \mbox{Binom}(\mathcal{I}_i^{(3)},1-\mbox{exp}(-\Delta t \cdot \gamma')) \\ 
    \gamma' &= \gamma \cdot k \\
    \mbox{FOI}_i (t) = \left( 1-\sum_{j\neq i} p_a \frac{M_{i,j}}{H_i}  \right) & \cdot \mbox{FOI'}_i (t) + \sum_{j\neq i}\left( p_a \frac{M_{i,j}}{H_i} \cdot \mbox{FOI'}_j (t) \right) \\
    \mbox{FOI'}_i (t) &= \beta_i(t) \frac{\mathcal{I}_i(t)^\alpha}{H_i} \\
    \beta_i (t) &= R_{i} (t) \cdot \gamma \\
    \mathcal{I}_i(t) &= \sum_{j=1}^{k=3} \mathcal{I}_i^{(j)}(t), 
\end{align}
where $M_{i,j}$ represent the daily mobility from location $i$ to location $j$, $H_i$ is the population of location $i$, $p_a$ is the proportion of time that moving individuals spend away, and $\alpha$ is the mixing coefficient. 

\section*{Conflict of Interest Statement}
The authors declare that the research was conducted in the absence of any commercial or financial relationships that could be construed as a potential conflict of interest.

\section*{Author Contributions}
VT and TM contributed to the conception and design of the study and to the data collection. VT implemented the system, performed data curation, conducted the experiments, and wrote the first draft of the manuscript. TM validated the progress and results of the study via daily discussion with VT. Both authors contributed to manuscript revision and read and approved the submitted version.

\section*{Funding}
This work was supported with funding from the COVID-19 Program and the Future Investment Program of the Research Organization of Information and Systems, Japan. 

\section*{Acknowledgments}
We are grateful to the members of the COVID-19 Project at our institute for their valuable discussions in frequent meetings. 

\section*{Data Availability Statement}
The data analyzed in this study were obtained from Twitter (tweet counts), JX Press (COVID-19 epidemic state), and ZENRIN DataCom (mobility data) and used in accordance with the licenses and restrictions of Twitter's ``Developer Agreement and Policy," JX Press' ``License for Research Purposes," and ZENRIN DataCom's ``License for Research Purposes." Requests to access these datasets should be directed respectively to \url{https://twitter.com/}, \url{https://jxpress.net/}, and \url{https://www.zenrin-datacom.net/}.

\bibliographystyle{frontiersinSCNS_ENG_HUMS} 
\bibliography{test}

\end{document}